\documentclass[12pt,letterpaper]{article}
\usepackage{osajnl}
\usepackage{xspace}
\newcommand{\ignore}[1]{}
\newcommand{\citep}[1]{\cite{#1}}
\newcommand{\cpu}{central processing unit (CPU)\renewcommand{\cpu}{CPU\xspace}\xspace}
\newcommand{\Ao}{Adaptive optics (AO)\renewcommand{\Ao}{AO\xspace}\renewcommand{\ao}{AO\xspace}\xspace}
\renewcommand{\ao}{adaptive optics (AO)\renewcommand{\ao}{AO\xspace}\renewcommand{\Ao}{AO\xspace}\xspace}
\newcommand{\elts}{extremely large telescopes (ELTs)\renewcommand{\elts}{ELTs\xspace}\renewcommand{\elt}{ELT\xspace}\xspace}
\newcommand{\elt}{extremely large telescope (ELT)\renewcommand{\elts}{ELTs\xspace}\renewcommand{\elt}{ELT\xspace}\xspace}
\newcommand{\mcao}{multi-conjugate AO (MCAO)\renewcommand{\mcao}{MCAO\xspace}\xspace}
\newcommand{\xao}{extreme AO (XAO)\renewcommand{\xao}{XAO\xspace}\xspace}

\newcommand{\fpga}{field programmable gate array (FPGA)\renewcommand{\fpga}{FPGA\xspace}\renewcommand{\fpgas}{FPGAs\xspace}\xspace}
\newcommand{\fpgas}{field programmable gate arrays (FPGAs)\renewcommand{\fpga}{FPGA\xspace}\renewcommand{\fpgas}{FPGAs\xspace}\xspace}
\newcommand{\hdl}{hardware description language (HDL)\renewcommand{\hdl}{HDL\xspace}\xspace}
\newcommand{\ffts}{fast Fourier transforms (FFTs)\renewcommand{\ffts}{FFTs\xspace}\renewcommand{\fft}{FFT\xspace}\xspace}
\newcommand{\fft}{fast Fourier transform (FFT)\renewcommand{\ffts}{FFTs\xspace}\renewcommand{\fft}{FFT\xspace}\xspace}
\newcommand{\shwfs}{Shack-Hartmann wavefront sensor (SHWFS)\renewcommand{\shwfs}{SHWFS\xspace}\xspace}
\newcommand{\psfs}{point spread functions (PSFs)\renewcommand{\psfs}{PSFs\xspace}\renewcommand{\psf}{PSF\xspace}\xspace}
\newcommand{\psf}{point spread function (PSF)\renewcommand{\psfs}{PSFs\xspace}\renewcommand{\psf}{PSF\xspace}\xspace}
\newcommand{\mpi}{message passing interface (MPI)\renewcommand{\mpi}{MPI\xspace}\xspace}
\newcommand{\Glao}{Ground layer \ao (GLAO)\renewcommand{\Glao}{GLAO\xspace}\renewcommand{\glao}{GLAO\xspace}\xspace}
\newcommand{\glao}{ground layer \ao (GLAO)\renewcommand{\Glao}{GLAO\xspace}\renewcommand{\glao}{GLAO\xspace}\xspace}
\newcommand{\wfs}{wavefront sensor (WFS)\renewcommand{\wfs}{WFS\xspace}\xspace}

 \newcommand{\mnras}{MNRAS}
 \newcommand{\aap}{A\&A}

\begin{document}
\title{Adaptive optics simulation performance improvements using reconfigurable logic}
\author{Alastair Basden}

\address{Centre for Advanced Instrumentation, Department of Physics, Durham
  University, South Road, Durham, DH1 3LE, UK}
\email{a.g.basden@durham.ac.uk}
\date{2006}

\begin{abstract}
A technique used to accelerate an adaptive optics simulation platform
using reconfigurable logic is described.  The performance of parts of
this simulation have been improved by up to 600 times (reducing
computation times by this factor) by implementing algorithms within
hardware and enables adaptive optics simulations to be carried out in
a reasonable timescale.  This demonstrates that it is possible to use
reconfigurable logic to accelerate computational codes by very large
factors when compared with conventional software approaches, and this
has relevance for many computationally intensive applications.  The
use of reconfigurable logic for high performance computing is
currently in its infancy and has never before been applied to this
field.
\end{abstract}

\section{Introduction}
The sensing of a corrupted optical wavefront is a key part of any
astronomical \ao system on an optical or infra-red telescope, and is
carried out using a \wfs, as described by Roddier\cite{roddier}.  When
starlight passes through the Earth's atmosphere, random perturbations
are introduced which distort the wavefronts from the astronomical
source in a time varying fashion \citep{tatarski}.  It is then no
longer possible to form a diffraction limited image from these
distorted wavefronts, and the effective resolution of a telescope is
reduced.  By sensing the form of the wavefront using a \wfs, and then
rapidly applying corrective measures to one or more deformable
mirrors, it is possible to compensate for some of the perturbations,
and hence improve the image quality and resolution of the telescope.
The \wfs and deformable mirror together form part of an \ao system.
\Ao is a technology widely used in optical and infra-red astronomy,
and almost all large science telescopes have an \ao system.  A large
number of results, which would be impossible to obtain using
seeing-limited (uncorrected) observations, have been obtained using
\ao systems (see for example Gendron\cite{2004A&A...417L..21G} and
Masciadri\cite{2005ApJ...625.1004M}).  However, there is still much
room for improvement: New \ao systems are continually being built and
new ideas developed, for example for wide-field high resolution
imaging \citep{2004SPIE.5490..236M} and extra-solar planet finding
\citep{2004ASPC..321...39M}.

The software simulation of an \ao system is an important part of the
characterisation of this \ao system.  This characterisation can be
used to determine whether a given \ao system will meet its design
requirements, thus allowing scientific goals to be met, or to model
new concepts.  The simulated performance of different \ao techniques
can be compared\cite{2005MNRAS.357L..26V}, allowing informed decisions
to be made when designing or upgrading an \ao system and when
optimising the system design.

A full \ao simulation will typically involve several stages
\citep{basden5}, from generation of simulated atmospheric phase
screens, image creation for wavefront sensing and system performance
categorisation, as well as simulation of the effect of the deformable
optical path elements and control algorithms.  The computational
requirements for \ao simulation scale rapidly with telescope size, and
simulation of \ao systems for the largest telescopes cannot be carried
out within acceptable timescales without the use of techniques to
greatly reduce computation time.

\subsection{Hardware acceleration}
The use of reconfigurable logic to provide application acceleration
for scientific applications is a relatively new area of research, and
I have used reprogrammable logic in the form of
\fpgas to provide hardware acceleration for \ao simulations.  An \fpga
is a user programmable logic array, which allows a hardware programmer
to link together the various elements within the \fpga in such a way
that enables the desired calculations to be carried out.  An \fpga can
perform highly parallelised calculations, carrying out simultaneous
independent operations in different parts of the device.  This high
degree of parallelism means that a large number of operations can be
performed simultaneously.

The clock speed of an \fpga is typically only a tenth of that of a
commodity computer processor.  However, due to the massively parallel
architecture, it is possible to program an \fpga so that a given
algorithm is computed at a much greater rate than would be possible
using a conventional \cpu.  In this paper, I show how reprogrammable
logic in the form of \fpgas has been used to greatly improve the
performance of a wavefront sensing algorithm, and has been integrated
with the Durham \ao simulation platform \citep{basden5}.

In \S2, I describe the wavefront sensing algorithm, and provide
details of the hardware implementation.  In \S3, I give results
for the performance improvements that are seen when using this
hardware acceleration.  In \S4, I describe future
work, and in \S5 I give our conclusions.

\section{The wavefront sensor pipeline}
In a real astronomical \ao system, starlight to be used for wavefront
sensing will be diverted from the main science beam using, for
example, a beam splitter.  This diverted beam will then usually be
passed through optical elements designed to allow the wavefront shape
to be sensed, such as a Shack-Hartmann lenslet array or a pyramid
optical element.  The beam is then imaged onto a detector, usually a
CCD, converted to an electronic form and passed to a processing
engine.  The processing engine will then use the detected light to
determine the shape of the wavefront, for example by employing a
centroiding algorithm in the case of a Shack-Hartmann system.  The
computed shape of the wavefront is then used to shape an optical
element, typically a deformable mirror, using a ``reconstruction''
process.  Designs for future \ao systems can require this process to
be repeated at a rate of 1-5~kHz with tens of thousands of degrees of
freedom in the reconstruction process\cite{2004SPIE.5490..195H}.  I now
consider only the case of a Shack-Hartmann wavefront sensing system,
and an overview of this wavefront sensing process is shown in
Fig.~\ref{fig:wfspipeline}.

\ignore{\begin{figure}
\includegraphics[width=8cm]{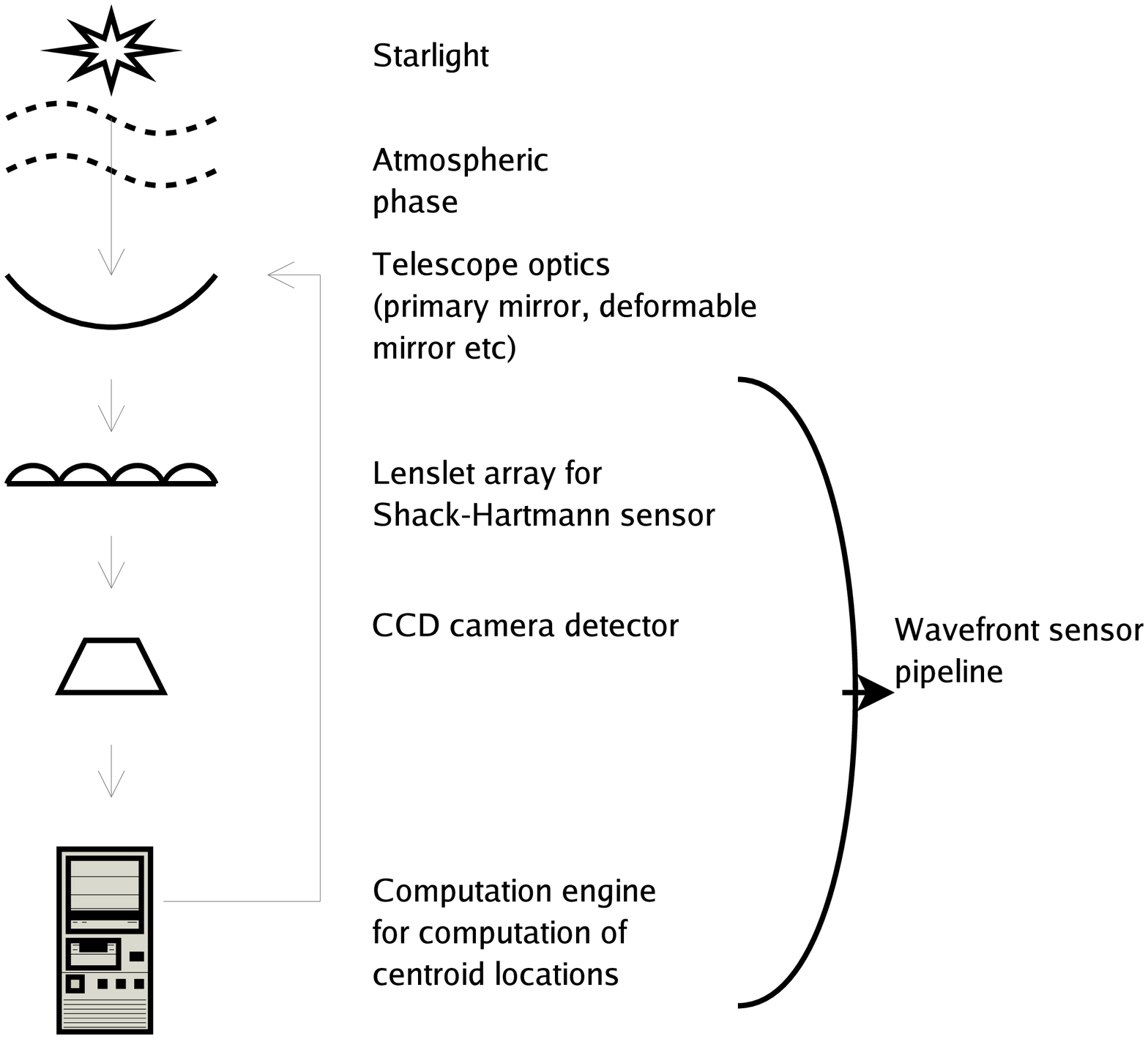}
\caption{A schematic diagram of the wavefront sensing process for a
  typical Shack-Hartmann wavefront sensor.}
\label{fig:wfspipeline}
\end{figure}}

\subsection{The simulated wavefront sensor pipeline}
\label{sect:pipecontents}
When simulating the wavefront sensing process, it is necessary to
model many physical processes as well as the computational processes
(such as the centroid location algorithm) so that the result will be
as accurate as possible.  When simulating a Shack-Hartmann \wfs with
the Durham \ao simulation platform, I start by computing the
atmospheric perturbations that have been introduced into the incident
starlight by the time this light has reached the telescope.  The
wavefront sensing process then computes the following steps:

\begin{enumerate}
\item A small phase map (equivalent to a phase tilt) is added to the
  atmospheric phase in each Shack-Hartmann sub-aperture, so that in
  the un-aberrated case (with no atmospheric perturbations) the
  maximum intensity will be placed at the centre of the central four
  pixels in the sub-aperture once the high light level Shack-Hartmann
  images are created.
\item A telescope pupil map is used to determine which parts of the
  telescope aperture are in the optical path.
\item The real and imaginary parts of the atmospheric phase array are
  computed (assuming an amplitude of unity) by taking the sine and
  cosine of the phase values.
\item The complex phase values are Fourier transformed using a two
  dimensional \fft.
\item The high light level (noiseless) image for each sub-aperture is
  computed by taking the square modulus of the Fourier transform.
\item Any required integration time and re-sampling of the image scale
  are performed on the high light level images.
\item The light within each sub-aperture is normalized so that the
  total light within the sub-aperture is equal to that expected from a
  given source magnitude.
\item A sky-brightness pattern is introduced into the high light level
  images.  
\item Photon shot noise is then added, by replacing the high light
  level (noiseless) image intensities in each pixel with a Poisson
  random variable with a mean and variance equal to this intensity.
\item The CCD readout noise is simulated by adding a Gaussian random
  variable with a mean and variance defined by the level of CCD
  readout noise to be simulated.  The simulated signal is now at the
  stage where it would be read out from the CCD camera and grabbed
  into computer memory.
\item Noise sources are subtracted from the signal, for example by
  applying a threshold value.
\item The centroid location of light within each sub-aperture is then
  computed.
\end{enumerate}

The centroid locations which are computed by this process are then
passed into a software wavefront reconstructor which will typically
use a large matrix multiplication operation to determine and update
the deformable mirror shape.

\subsection{Hardware implementation of the wavefront sensing pipeline}
When implementing an algorithm in an \fpga, it is important to
consider how the algorithm can be parallelised so that the \fpga is
used efficiently.  I have implemented the \wfs pipeline
in a way which means that all stages of the pipeline can operate
simultaneously on different parts of a dataset.  This is demonstrated in
Fig.~\ref{fig:pipeline}.  In simple terms, the phase data for one
sub-aperture is loaded into the \fpga.  The \fpga then begins to
compute the 2D Fourier transform of this data while a second dataset
is loading into the \fpga.  After the Fourier transform has been
computed, various noise sources (e.g.\ photon shot noise) are
introduced into the sub-aperture.  While this is happening, the
second dataset is being Fourier transformed, and the third dataset is
being loaded into the \fpga.  This highly parallel operation then
continues until the results for all sub-apertures have been computed.
In reality, the parallelization is even finer as, for example, the
Fourier transform will begin while data is still being loaded, and the
introduction of photon shot noise will be computed in stages on many
different pixel values at the same time.

\ignore{\begin{figure}
\includegraphics[width=8cm]{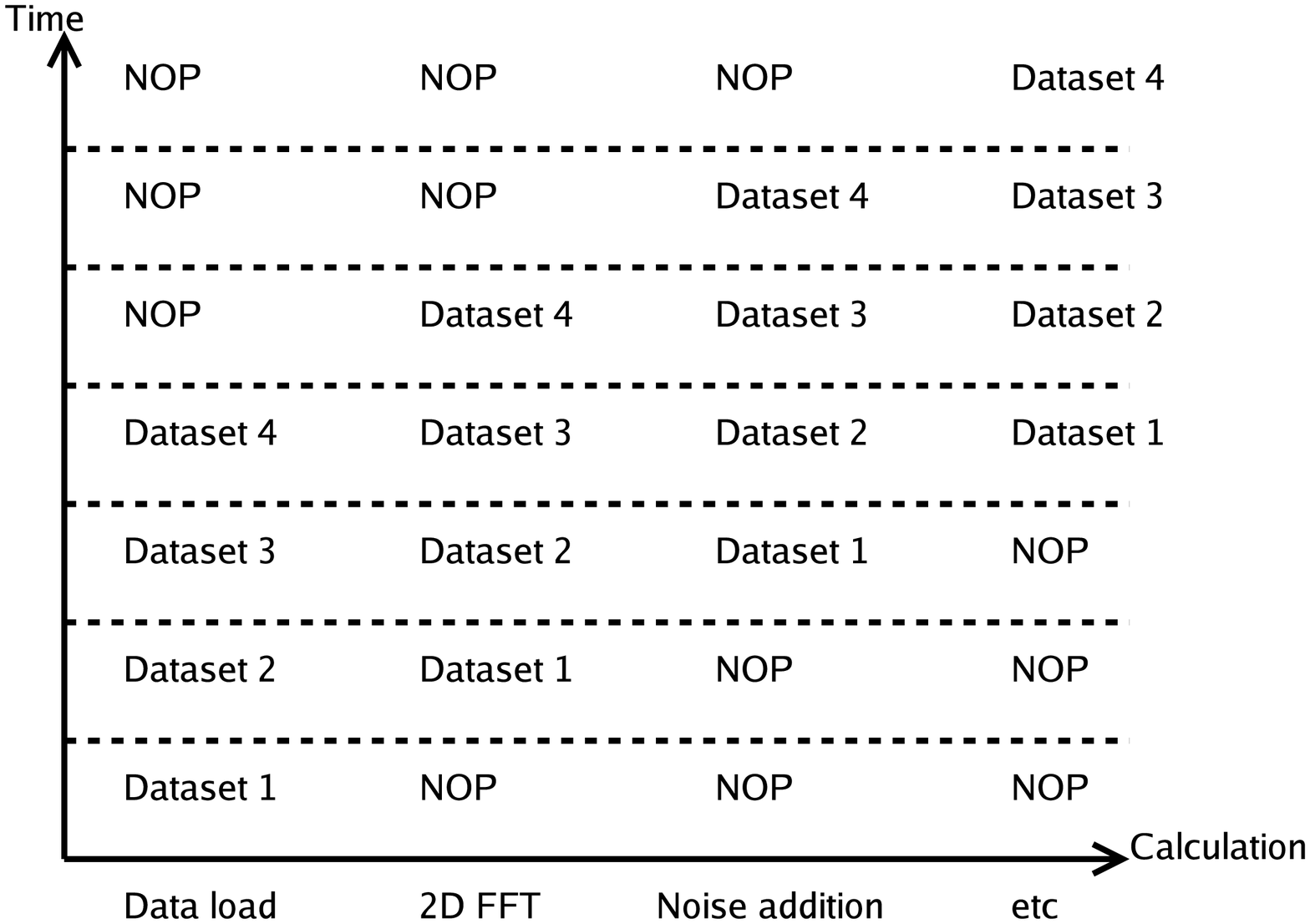}
\caption{A schematic diagram showing the parallelization of the
  wavefront sensor pipeline within the FPGA.  All stages operate
  simultaneously.  Here, NOP means no operation is computed, i.e.\ the
  input data is not valid.}
\label{fig:pipeline}
\end{figure}}

In a \cpu implementation only one stage can happen at a time and so
even though the time to compute each stage may be less than with the
\fpga implementation, the total computation time will be greater.

\subsubsection{The algorithm choice}
Many simulation codes (for example propagation codes) contain one
part of the simulation where the majority of the computation time is
spent computing a simple algorithm.  Such simulations are ideal
candidates for hardware acceleration because the simple algorithm can
easily be placed in hardware, and the resulting performance increase
of this part of the simulation also gives a similar performance
increase to the simulation as a whole, since this is where the
majority of computation time is spent.

However, in an \ao simulation, a large amount of computational time is
divided between a number of key components since an \ao simulation
contains many complex algorithms, for example, the reconstruction
algorithms, the atmospheric phase screen generation, and various parts
of the wavefront sensing pipeline.  To give a large overall
performance increase for an \ao simulation, each of these components
would require a performance increase.  Amdahl's law \citep{amdahl}
states that the overall system speed is governed by the slowest
component.  As an example, a simulation may consist of five
algorithms, each requiring 20~percent of processor time.  If one of
these algorithms was then implemented in hardware, which reduced the
computational time for this algorithm by a factor of 1000, the overall
simulation computational time will only be reduced by about a fifth.
It is therefore necessary to reduce the computational time of all five
algorithms to achieve an overall performance increase of greater than
five times.

This means that most parts of the \ao simulation should be implemented
in hardware to give a significant performance increase.  I have
therefore implemented the wavefront sensing algorithms in hardware as
a first step, since these algorithms are among the most
computationally intensive.  Additionally, these algorithms are among
the most complicated to place in hardware and so success implementing
them gives an idea of the difficulty of implementing the majority of
an \ao simulation within one or more \fpga.

\subsubsection{Data bandwidth considerations}
In many computer architectures, data flow -- passing data between
computational elements -- can cause a bottleneck when processes spend
significant amounts of time waiting for data.  On the Cray XD1 super
computer where the hardware wavefront sensing pipeline has been
implemented, it is possible to pass data at a theoretical bandwidth of
1.6~GBs$^{-1}$ between the \fpga and processor memory in each
direction.  The highly parallel nature of the \fpga means that many
calculations can be computed simultaneously.  It is therefore
important to ensure that all of these calculations can be fed with
necessary data.  To do this, I have ensured that data is passed to
and from the \fpga only once, with no intermediate results being
returned to the \cpu memory.  This ensures that the \fpga can be fed
with new data at all times.  If, for example, one intermediate result
stage was returned to the \cpu memory, operated on by the \cpu, and
then passed back into the \fpga, the bandwidth available for each data
transfer stage would be halved, resulting in an increase in the
computation time.

It is therefore essential to program the \fpga so that data flow to and
from the host processor memory is minimised.

\subsubsection{Algorithms within the wavefront sensing pipeline}
Optical phase data (aberrated by the atmospheric turbulence) is
currently generated by the \cpu, and read by the \fpga at every
iteration (time-step) of the simulation.  Once computed within the
\fpga, the floating point centroid locations are written back to the
\cpu main memory by the \fpga.  I have implemented all of the
algorithms described in section~\ref{sect:pipecontents} within the
hardware implementation of the wavefront sensing pipeline, which is
treated as a black box, accepting floating point optical phase data,
and returning floating point centroid location values.  Internally,
data is stored in the most appropriate format chosen to give the
required precision while not consuming \fpga resources which are not
needed.  For example, during the high light level image computation,
data is stored in a floating point format with a 22 bit mantissa and a
six bit exponent, while during the introduction of photon shot noise, the
data is stored as a 23 bit wide fixed point number.  

It is important to realise that each stage within the pipeline can
operate simultaneously with other stages on different parts of the
dataset.  In total, about four months of effort was required to
implement this algorithm in hardware.  The software version on the
other hand could be prepared in about a week or so, giving some idea of the
differences between software and hardware complexities.  The hardware
algorithm was implemented in VHDL, a low level hardware description
language.  Higher level languages are available for hardware
programming (such as Handel-C).  However, these higher level languages
take up much more of the \fpga resources (typically by a factor of two
or much more), and can often restrict the maximum clock speed of the
\fpga for which the algorithm will give correct results.  These
factors mean that the wavefront sensing pipeline would not fit into a
single \fpga if created using a higher level language.  Currently,
about 70~percent of the \fpga resources (a Virtex-II Pro V2P50) are
used for this pipeline.

\subsubsection{Configuration}
\fpgas are usually considered to be fixed function devices, performing
only one set task (which may be simple or complicated).  In the case
described here, the fixed function is the \wfs algorithms.  However,
these have been implemented to be configurable:
\begin{enumerate}
\item The dimensions of the input optical phase array for each
sub-aperture can be selected, up to a maximum of $32\times 32$ values.
\item The size of a two dimensional \fft (used to create the high light
level images from the atmospheric phase) can be chosen from $8\times
8$, $16\times 16$ or $32\times 32$ pixels.  If the \fft dimensions are
not equal to the optical phase array dimensions, the optical phase is
zero-padded inside the \fpga before the \fft is executed.  
\item The number of iterations over which the wavefront sensor is to
be integrated before being read out is also configurable (up to a
maximum of 63 iterations).  
\item Pixel re-sampling (binning) can also be configured, with the
\fft output being re-sampled to any size from $2\times2$ pixels up to
the size of the \fft (including rectangular arrays).  
\item Random number generator seeds can be configured.
\item The shape of the telescope pupil function which is used to
determine which sub-apertures are able to collect light can be defined.
\item CCD readout noise parameters can be configured (mean and
root-mean-square).
\item A sky background value can be set.
\item A threshold level to be applied to the signal after CCD readout
has been simulated can be set.
\item The number of sub-apertures is configurable up to a maximum of
$1024\times1024$ (more sub-apertures can be used when a pupil mask is
  not required).
\item The magnitude of the guide star can be chosen.
\end{enumerate}

The configurability of this hardware implementation of the wavefront
sensing and centroid algorithms means that the benefits of hardware
acceleration can be realized for a large range of simulations.

\subsection{Integration with the AO simulation platform}
The hardware accelerated wavefront sensing algorithms have been
integrated with the Durham \ao simulation platform in such a way that
the user needs to specify only whether or not the hardware
implementation should be used where possible.  There are situations in
which this implementation cannot be used (for example when simulating
a Pyramid \wfs), in which case, a software algorithm is
used instead.  When an \ao simulation is running, it is possible to
switch on or off the \fpga acceleration facility using a graphical
simulation control interface.

\section{Performance improvements}
An \fpga should be operated at a clock rate which is dependent on the
logic implemented within the \fpga.  The synthesis tools used to
compile the \fpga code give an indication of what the maximum clock
rate should be.  Improving the implementation of the logic within the
\fpga will often allow a faster clock rate to be used, for example by
more efficient pipelining.  When the clock rate is set too high, the
\fpga will cease to function in the expected way, which can cause data
corruption.  The \fpga performance is directly dependent on this clock
rate.

\subsection{FPGA clock rate performance}
The relative time taken by the software and hardware algorithms
determines the performance increase achieved by the hardware
implementation.  For this purpose, a slightly modified software
algorithm was used which computed only the algorithms also computed in
the \fpga.  Fig.~\ref{fig:speedup} shows the performance increase
obtained as a function of \fpga clock speed, and it can be seen that
the trend is linear.  The dotted line in Fig.~\ref{fig:speedup} shows
that at clock speeds above about 170~MHz, the hardware implementation
of the algorithm becomes unreliable due to overclocking of the \fpga,
sometimes giving incorrect results.  Further work on the algorithms
within the \fpga to streamline the pipeline would allow the \fpga to
give accurate results at these clock rates, though this has not yet
been carried out.  In the Cray XD1, the \fpgas can be clocked at a
maximum rate of 199~MHz, which would give an extra 10~percent increase
in performance over the 170~MHz clock rate which currently gives
correct results.  For the rest of this paper, a clock rate of 170~MHz
is assumed unless otherwise stated, so that data integrity is
maintained.

\ignore{\begin{figure}
\includegraphics[width=8cm]{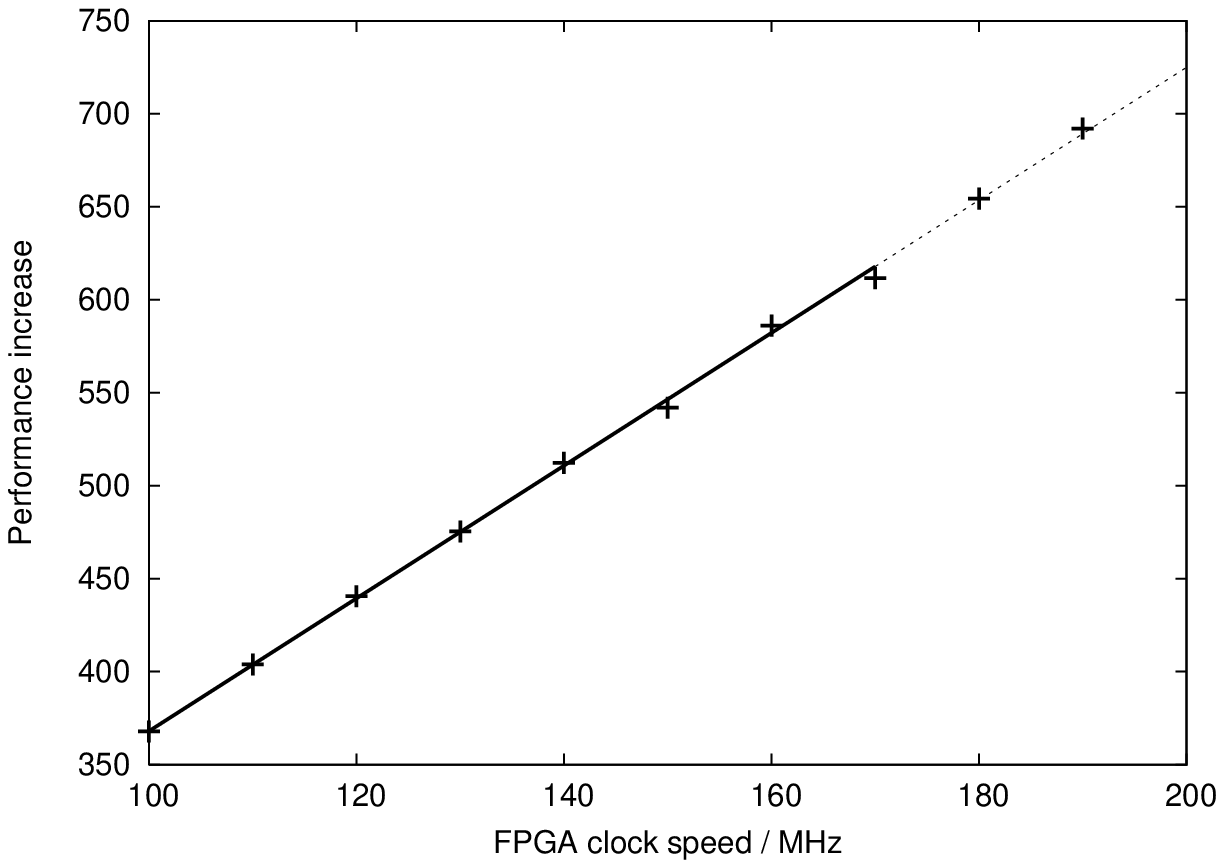}
\caption{A figure showing the performance increase when using an FPGA
  instead of a CPU for wavefront sensing pipeline algorithms, as a
  function of the FPGA clock rate.  The dotted line above 170~MHz
  shows the speeds at which the \fpga algorithm becomes unreliable due
  to overclocking.  Here, the centroid locations of 1024
  Shack-Hartmann sub-apertures have been computed at each \fpga clock
  frequency.}
\label{fig:speedup}
\end{figure}}

\subsection{Data quantity}
The relative performance improvement achieved when using the \fpgas
depends partly on the size of the dataset which is accessed by the
\fpga.  When the dataset is small, the overhead of reading data into
the \fpga, and writing the results back to the host \cpu memory will
be large compared to the time spent computing the results.  Therefore,
the performance improvement will be small (indeed, there are
algorithms in which the \fpga can be slower than the \cpu when the
dataset is small\cite{basden4}).  The pipeline itself also has some
latency, as there is a finite time between the last data entering the
pipeline, and the last result leaving the pipeline, of order 700 clock
cycles, or 4~$\mu$s (at 170~MHz).  Fig.~\ref{fig:nsubaps} shows the
time taken to compute the \wfs algorithms when using the
hardware and software implementations, as a function of the number of
sub-apertures operated on.  As can be seen, when using the \fpga
implementation, the total computation time is about 20~$\mu$s for
small numbers of sub-apertures (up to about $4\times4$), regardless of the
number of sub-apertures used, corresponding to the latency in the
pipeline and the latency for memory access.  When larger numbers of
sub-apertures are used, the computation time is proportional to the
number of sub-apertures evaluated. 

The time taken for the hardware \wfs algorithms to complete can be
estimated when the number of sub-apertures is large (greater than
about 100), as
\begin{equation}
t=\frac{n_f^2\times n_s \times n_i}{f} + t_l
\label{eq:comptime}
\end{equation}
where $t$ is the time taken in seconds, $n_f$ is the size of the
dimensions of the 2D \fft used to compute the high light level images
(8, 16 or 32), $n_s$ is the total number of sub-apertures to be
evaluated, $n_i$ is the number of integrations carried out, $f$ is
the clock frequency of the \fpga in Hz and $t_l$ is the initial
latency of the pipeline, about $20\times10^{-6}$~s.  The massively parallel
architecture means that the computation time is simply the computation
time of the lowest algorithm (in this case, creating and integrating
the high light level images).  

\ignore{\begin{figure}
\includegraphics[width=8cm]{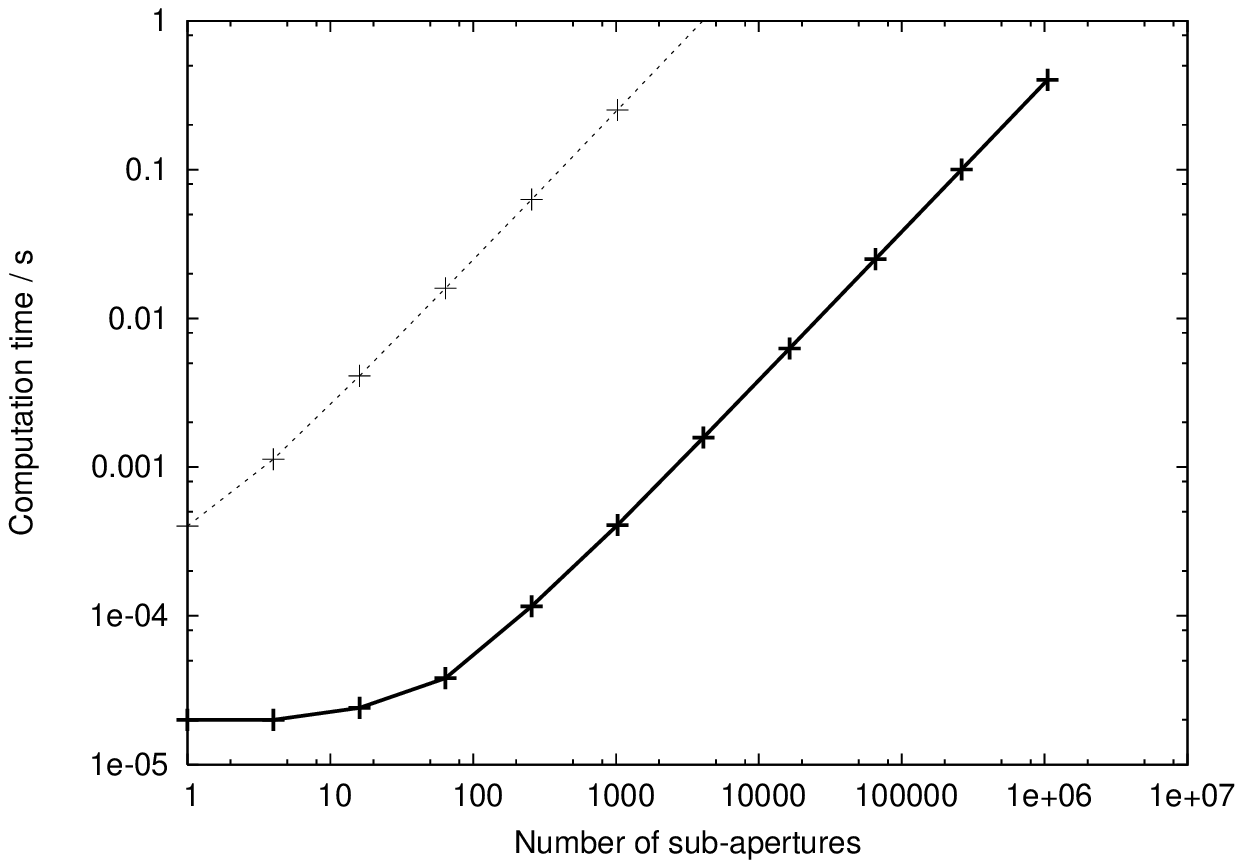}
\caption{A figure showing the computation time for the wavefront
  sensing pipeline in hardware (solid curve) and software (dotted
  curve) as a function of the number of sub-apertures to be evaluated.
  }
\label{fig:nsubaps}
\end{figure}}
When the \cpu implementation is used, the computation time is
dependent on the number of sub-apertures evaluated.  The memory access
latency is lower when the calculation is carried out in the \cpu, and
so the computation time scales closely with the number of
sub-apertures to be evaluated even when this is small.  The
performance increase when using the \fpga is therefore less when
smaller numbers of sub-apertures are used, as shown in
Fig.~\ref{fig:nsubapsratio}.  A typical current \ao system will
contain 100 sub-apertures, while future \ao systems are planned with
over 100,~000 sub-apertures when multiple wavefront sensors are used.

\ignore{\begin{figure}
\includegraphics[width=8cm]{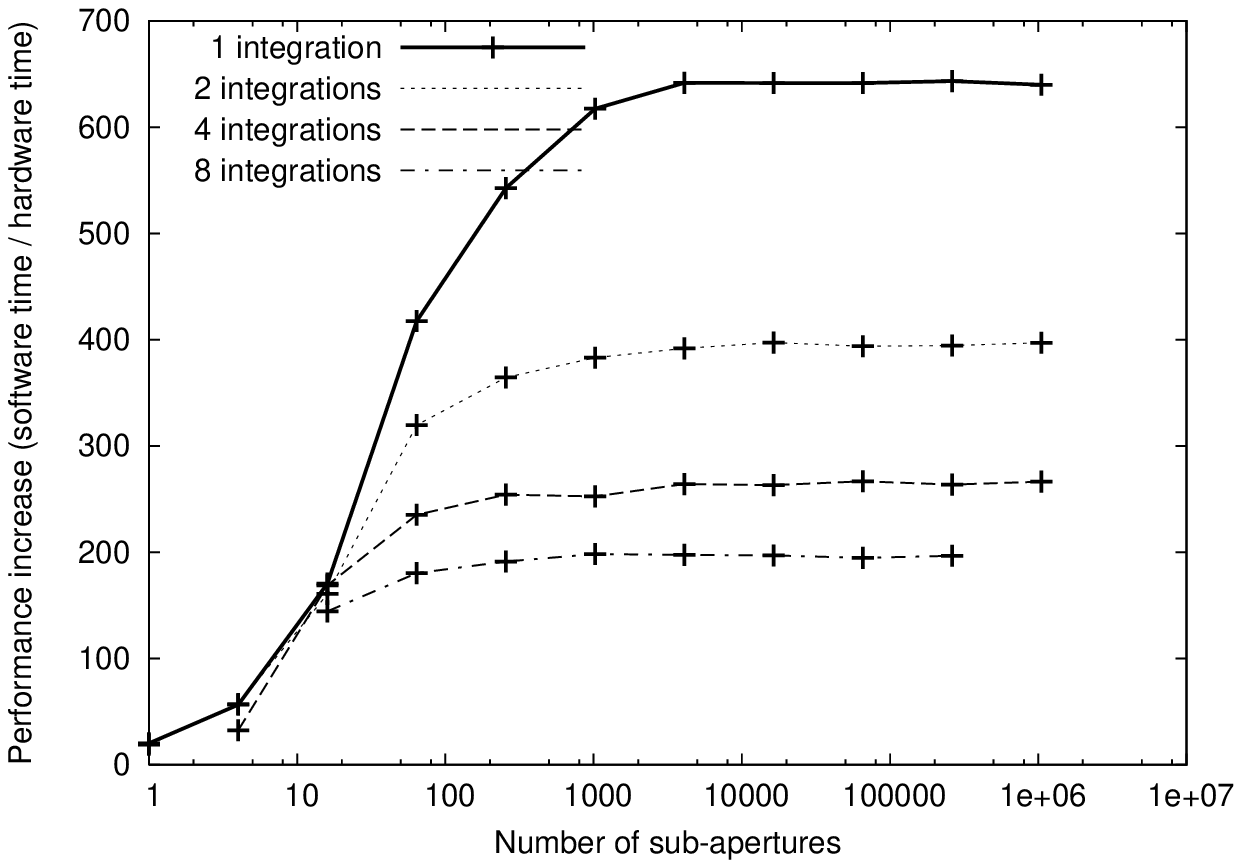}
\caption{A figure showing the ratio of CPU to FPGA computation time
  for the wavefront sensing pipeline as a function of the number of
  sub-apertures to be evaluated.  A typical current AO system will
  contain 100 sub-apertures, while future AO systems are designed with
  over 100,000 sub-apertures.  The number of image integrations
  carried out before the CCD readout is simulated is 1 (solid curve),
  2 (dotted curve), 4 (dashed curve) and 8 (dot-dashed curve).}
\label{fig:nsubapsratio}
\end{figure}}

\subsection{Integrations}
When the optical phase data is sampled more than once for each CCD
readout simulation, (i.e.\ the sub-aperture images are integrated
before photon shot noise or CCD readout noise is added), it is
necessary to pass more data into the \fpga per final centroid value.
Since the \fpga computation is limited by the rate at which data is
passed in, this will increase the computation time proportionally to the
number of integrations.  However, the time taken by the software
implementation will only be proportional to the number of integrations
up to the point at which the integration is carried out.  After this,
the remainder of the calculation (noise addition, centroid estimation)
will be performed only once, meaning the time taken is independent of
the number of integrations.  Therefore, the total time taken by the
software implementation will be less than proportional to the number
of integrations, meaning that the relative performance improvement
realized by the \fpga will be reduced as shown in
Fig.~\ref{fig:nsubapsratio}.

\subsection{Sub-aperture size}
The computation time of the \fpga implementation of these algorithms
is given by Eq.~\ref{eq:comptime}.  For the software implementation,
this is not the case, since an increase in sub-aperture dimensions by
a factor of two (i.e.\ four times as many phase value array elements
per sub-aperture) will take less than four times as long to compute,
as demonstrated in Fig.~\ref{fig:subapsize}, due to the different
rates at which various algorithms within the pipeline take to
complete.  This figure shows that performance increase provided by the
\fpga is reduced as the sub-aperture size increases.  Additionally,
whereas for the \fpga implementation, pixel re-sampling does not
affect the computation time (due to the fine grain parallel
architecture), the \cpu implementation run time is effected, taking
shorter times to complete when greater binning is used (due to the
centroid algorithm then being applied to smaller sub-apertures).  The
performance increase will therefore be less when the \cpu
implementation performs relatively faster as shown in
Fig.~\ref{fig:subapsize}.  In most \ao simulations, small
sub-apertures are used with typically $8\times 8$ phase array
elements.

\ignore{\begin{figure}
\includegraphics[width=8cm]{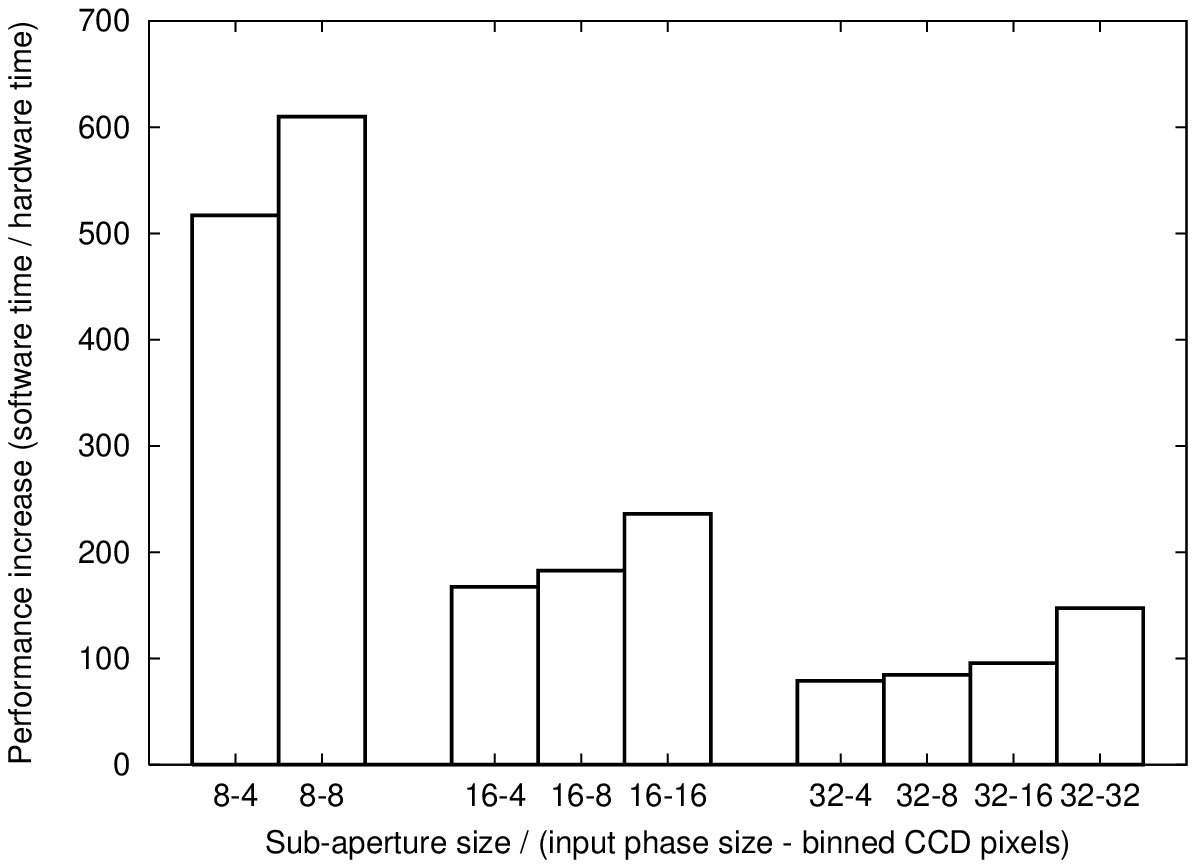}
\caption{A figure showing the ratio of CPU to FPGA computation time
  for the wavefront sensing pipeline as a function of the number of
  pupil phase values per sub-aperture (first number for each bar) and
  the number of simulated CCD pixels (in each dimension) per
  sub-aperture (second number for each bar).  }
\label{fig:subapsize}
\end{figure}}

\section{Future work}
The hardware \wfs pipeline is useful for a wide range of simulations.
The effect of spot elongation when using laser guide stars is not yet
considered in this pipeline, and so this could be added, Additionally,
taking scintillation effects into account may be necessary for some
systems.

Implementing these algorithms within the current \fpgas in the XD1 is
unlikely to be possible, due to the finite amount of logic within the
\fpgas.  However, it is possible to upgrade the XD1 with larger
\fpgas, and this would allow these extra algorithms to be implemented
within a single \fpga should funds become available.

In a typical software \ao simulation at Durham, the \wfs
algorithms may take about 75~percent of \cpu time.  By implementing
these algorithms in an \fpga, they can be accelerated by over 600
times.  The remaining 25~percent of the processor tasks then have full
access to the conventional processors.  However, this will then only
accelerate the \ao simulation by a factor of four, not particularly
impressive given the acceleration achieved for the \wfs
algorithms.  It is therefore necessary to implement other parts of the
simulation within the \fpgas.

The majority of the remaining processor tasks are found within the
reconstruction algorithms which map the \wfs outputs onto
new figures for the deformable mirror optics.  It is therefore
necessary to implement these algorithms in hardware to accelerate the
simulation further.  Given the large number of combinations of natural
guide stars, laser guide stars and conjugate deformable mirrors, it is
not possible to include the reconstruction algorithms as part of the
\wfs pipeline, but rather, in a separate hardware
implementation running within a different \fpga.  This will allow for
the flexibility and reconfigurability of the \ao simulation to be
maintained.

The construction of atmospheric turbulence phase screens is also
processor intensive, and it is planned to move this algorithm into
hardware also.  This again will improve the overall performance of the
\ao simulation.  

\section{Conclusion}
I have described the implementation of a \wfs simulation pipeline
within reconfigurable logic in the form of \fpgas, and this is the
first time that this has been attempted.  This has led to a reduction
in computation time of over 600 times over the conventional software
approach, allowing the simulation run times to be reduced.  This work
demonstrates the feasibility and benefits of hardware acceleration
using \fpgas, and shows that hardware acceleration can greatly improve
the performance of calculations, far beyond that achievable by using
software based approaches.  This has relevance for simulation of a
wide range of optical systems.  By using a hardware accelerated \ao
simulation platform, it is possible to model \ao systems on extremely
large telescopes, which would be otherwise infeasible due to the long
computation times.

\section*{Acknowledgments}
The author would like to thank R.~Wilson, C.~Saunter and D.~Geng for
their thoughtful comments.


\pagebreak
\begin{figure}[h]
\includegraphics[width=8cm]{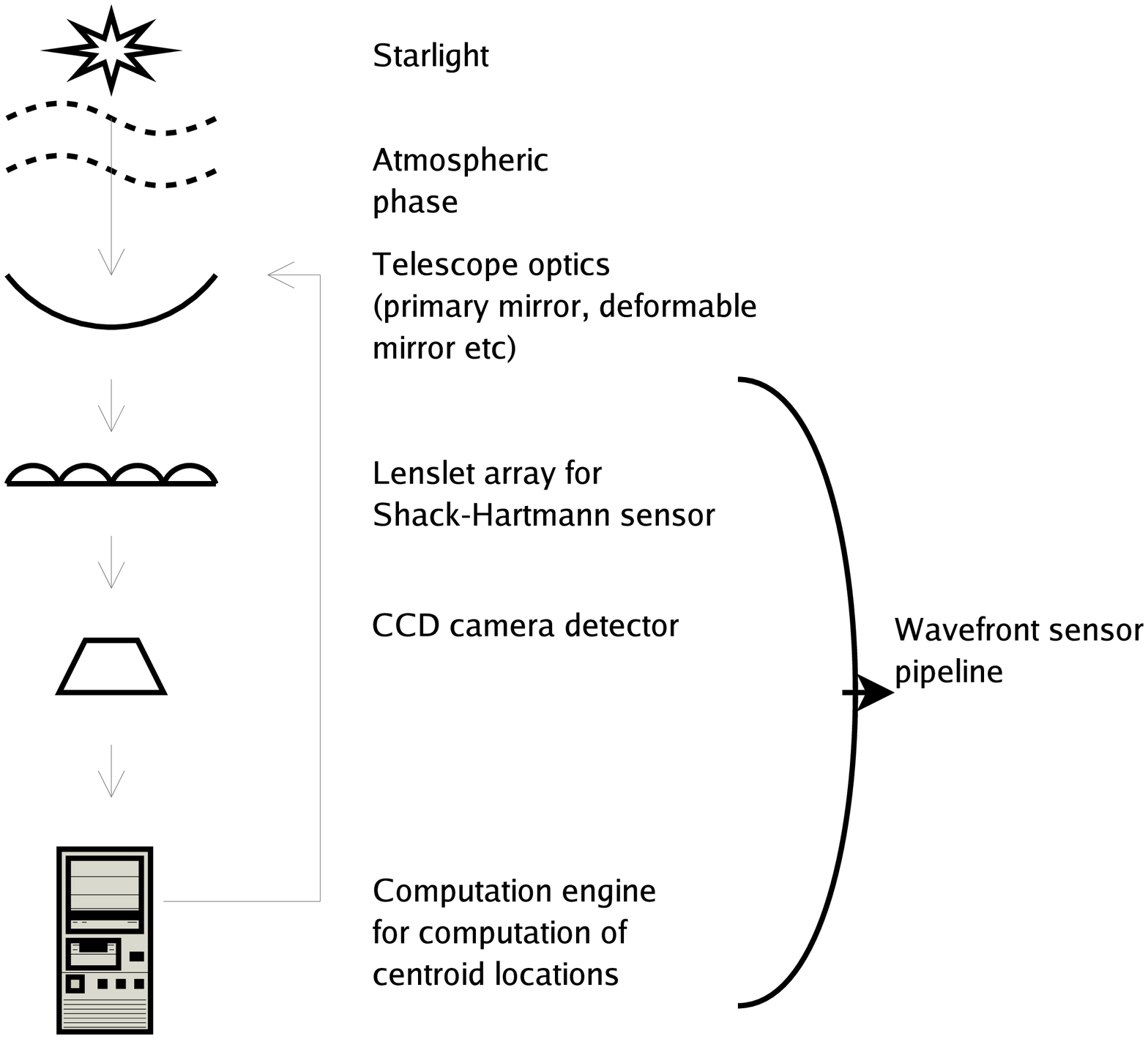}
\caption{A schematic diagram of the wavefront sensing process for a
  typical Shack-Hartmann wavefront sensor.}
\label{fig:wfspipeline}
\end{figure}

\pagebreak
\begin{figure}[h]
\includegraphics[width=8cm]{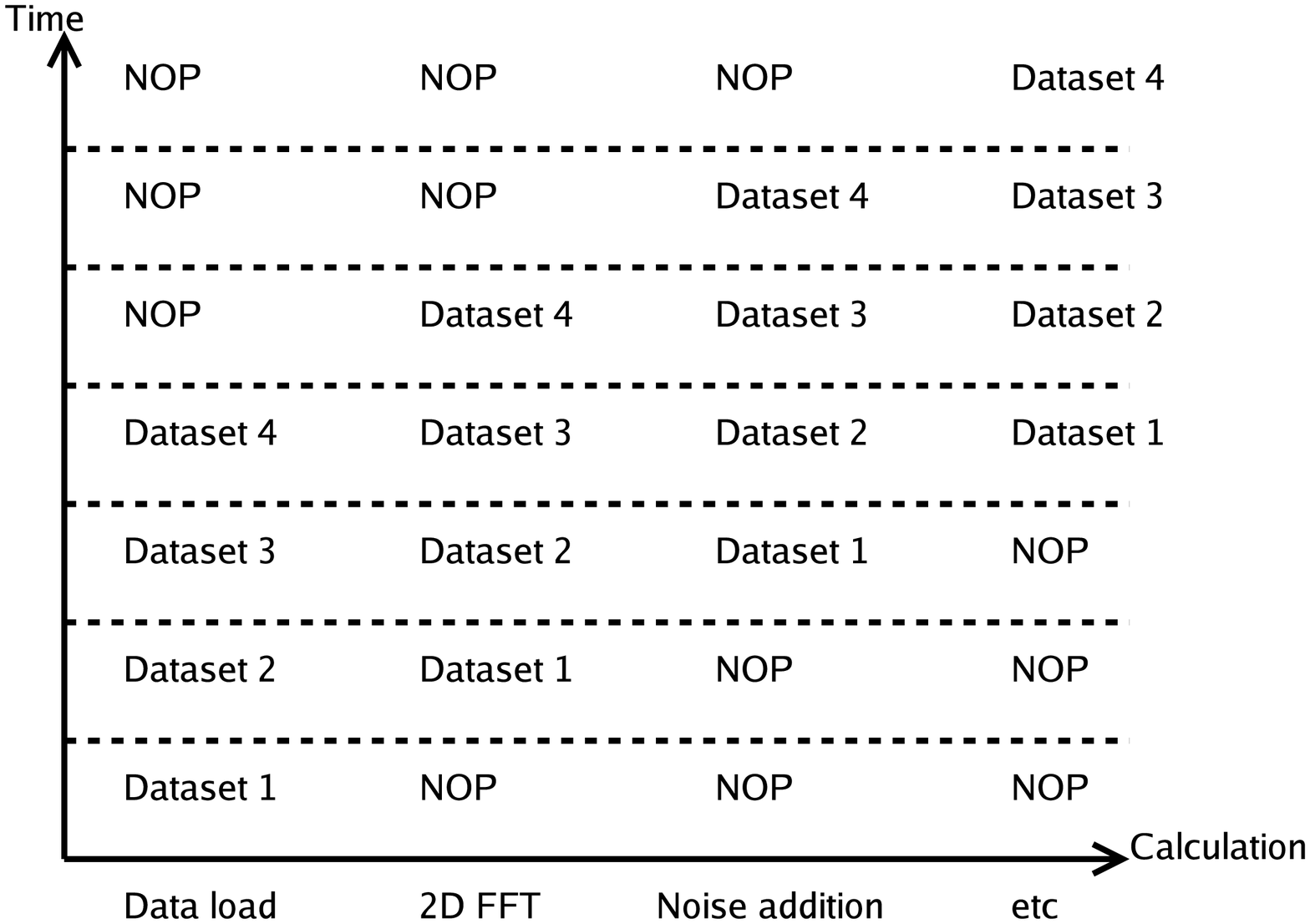}
\caption{A schematic diagram showing the parallelization of the
  wavefront sensor pipeline within the FPGA.  All stages operate
  simultaneously.  Here, NOP means no operation is computed, i.e.\ the
  input data is not valid.}
\label{fig:pipeline}
\end{figure}

\pagebreak
\begin{figure}[h]
\includegraphics[width=8cm]{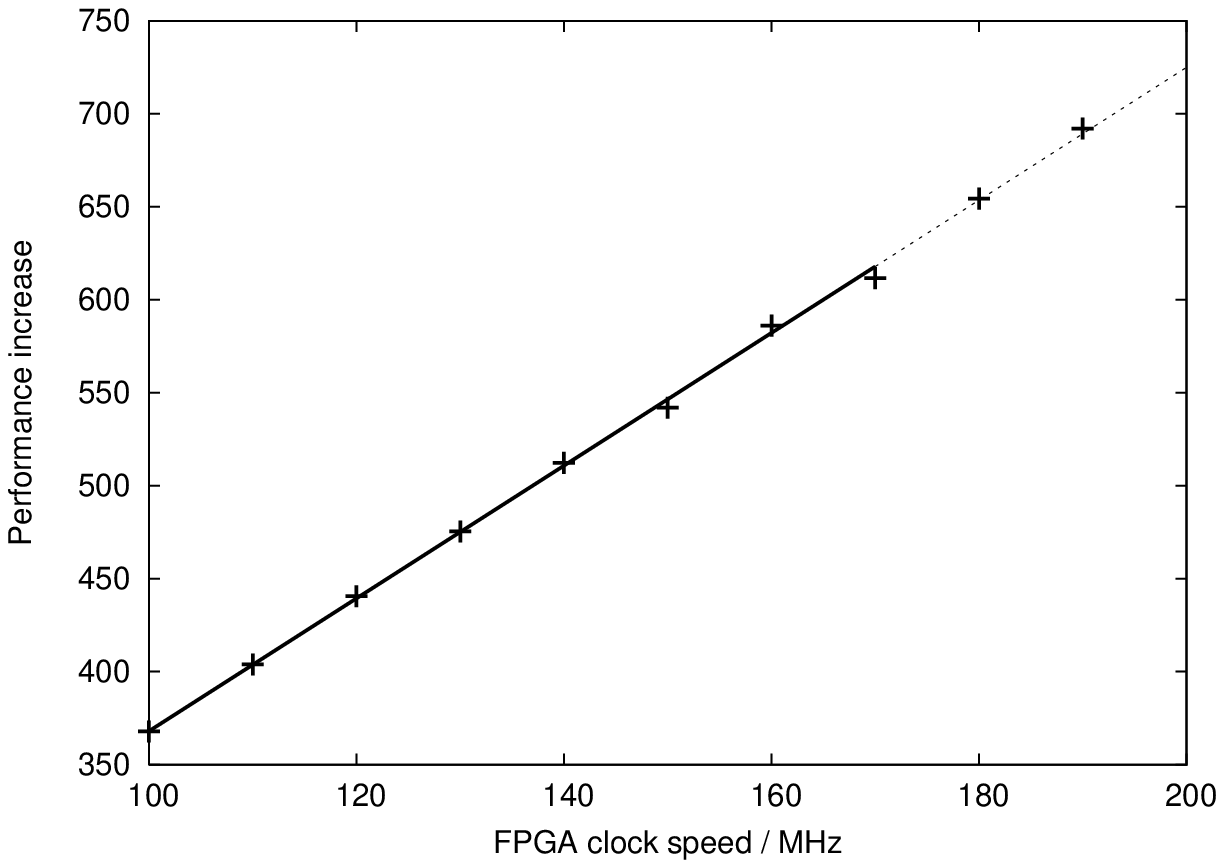}
\caption{A figure showing the performance increase when using an FPGA
  instead of a CPU for wavefront sensing pipeline algorithms, as a
  function of the FPGA clock rate.  The dotted line above 170~MHz
  shows the speeds at which the \fpga algorithm becomes unreliable due
  to overclocking.  Here, the centroid locations of 1024
  Shack-Hartmann sub-apertures have been computed at each \fpga clock
  frequency.}
\label{fig:speedup}
\end{figure}

\pagebreak
\begin{figure}[h]
\includegraphics[width=8cm]{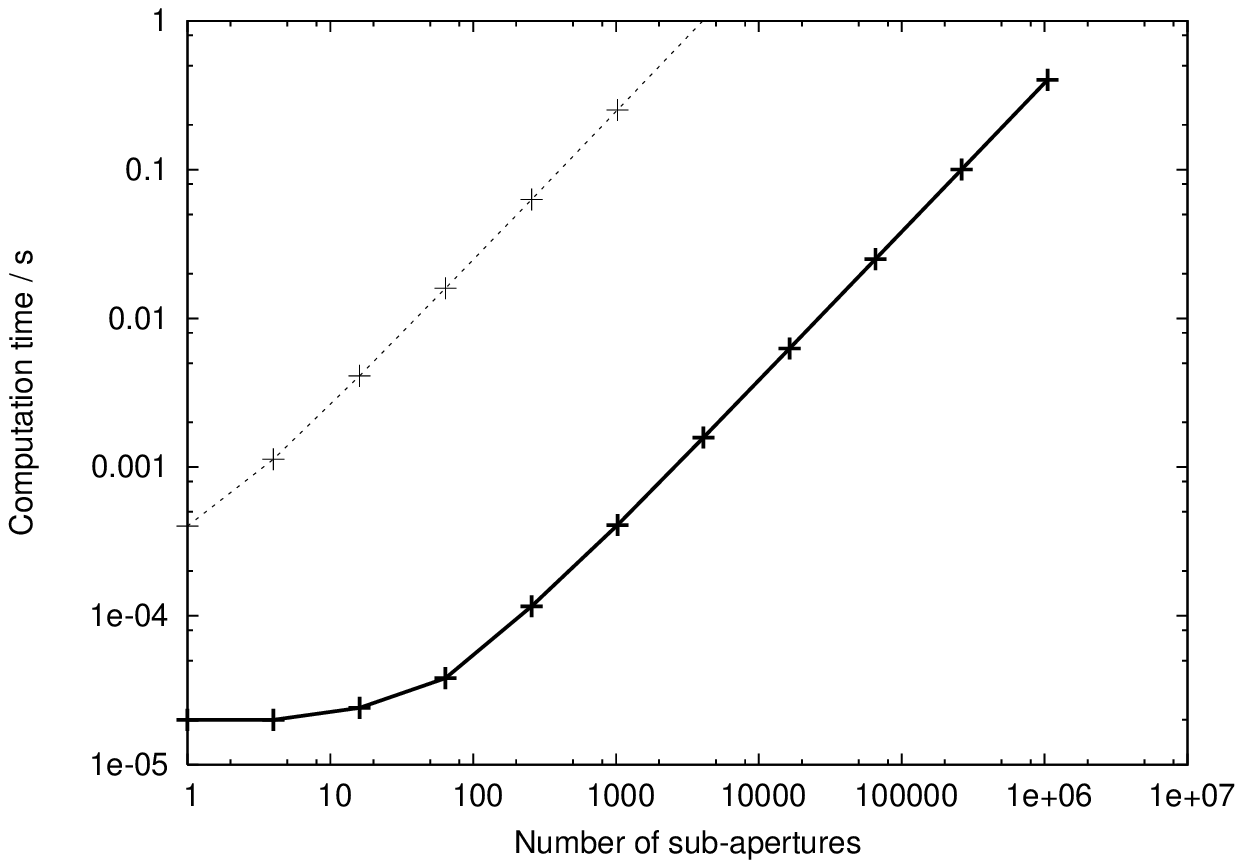}
\caption{A figure showing the computation time for the wavefront
  sensing pipeline in hardware (solid curve) and software (dotted
  curve) as a function of the number of sub-apertures to be evaluated.
  }
\label{fig:nsubaps}
\end{figure}

\pagebreak
\begin{figure}[h]
\includegraphics[width=8cm]{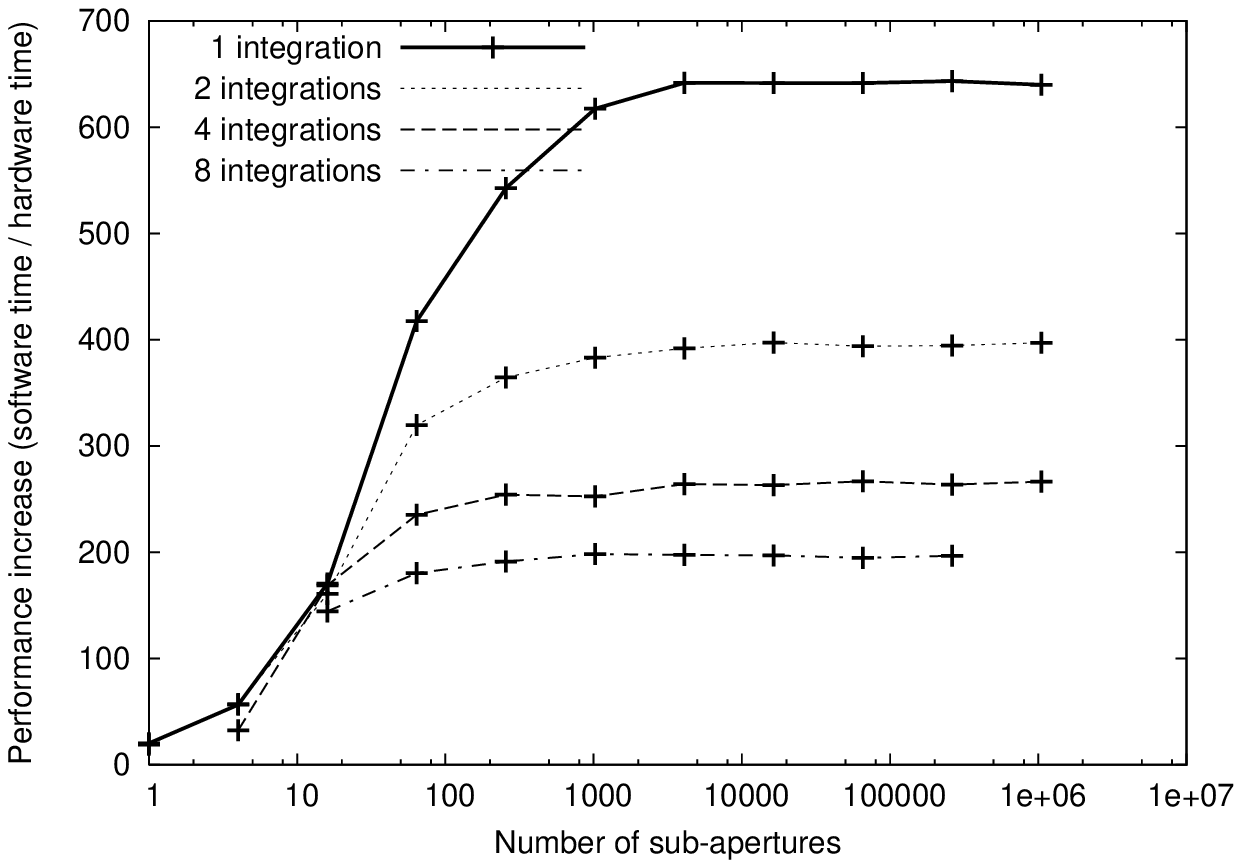}
\caption{A figure showing the ratio of CPU to FPGA computation time
  for the wavefront sensing pipeline as a function of the number of
  sub-apertures to be evaluated.  A typical current AO system will
  contain 100 sub-apertures, while future AO systems are designed with
  over 100,000 sub-apertures.  The number of image integrations
  carried out before the CCD readout is simulated is 1 (solid curve),
  2 (dotted curve), 4 (dashed curve) and 8 (dot-dashed curve).}
\label{fig:nsubapsratio}
\end{figure}

\pagebreak
\begin{figure}[h]
\includegraphics[width=8cm]{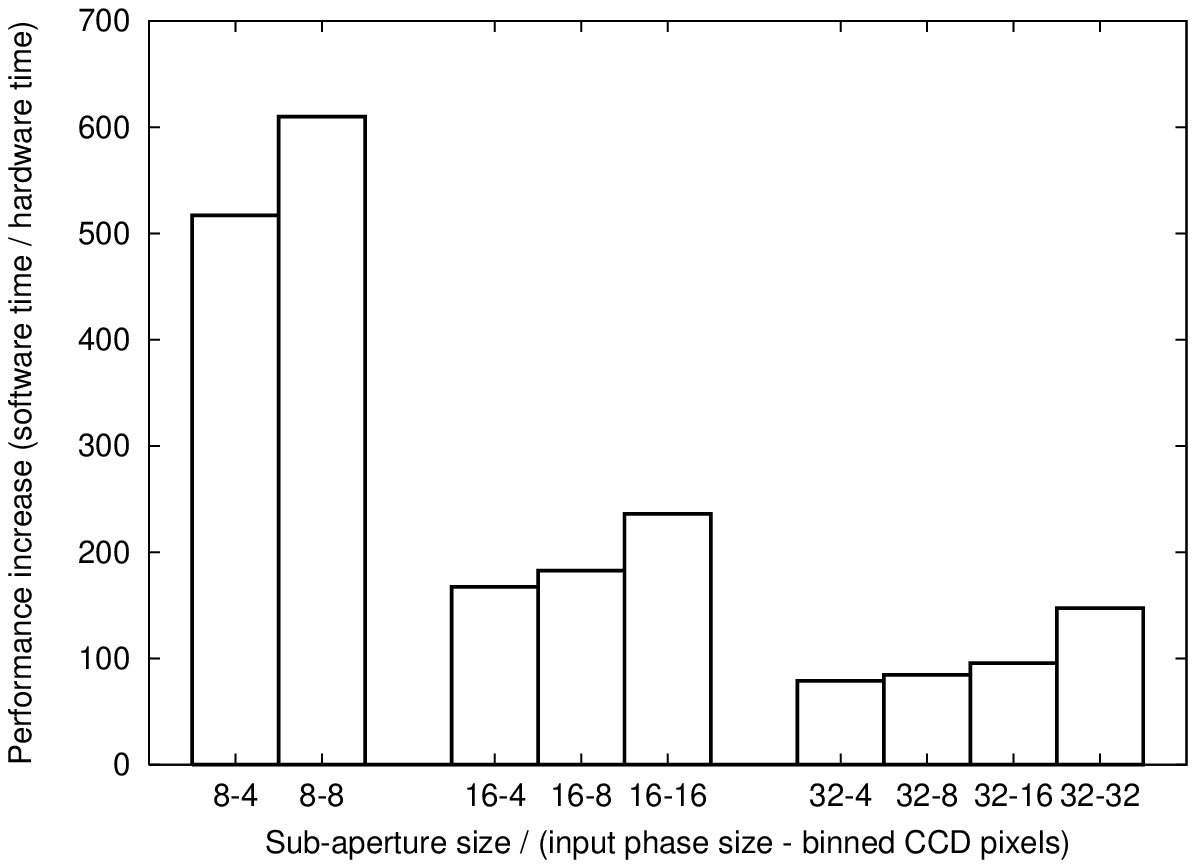}
\caption{A figure showing the ratio of CPU to FPGA computation time
  for the wavefront sensing pipeline as a function of the number of
  pupil phase values per sub-aperture (first number for each bar) and
  the number of simulated CCD pixels (in each dimension) per
  sub-aperture (second number for each bar).  }
\label{fig:subapsize}
\end{figure}

\pagebreak

\begin{enumerate}
\item A schematic diagram of the wavefront sensing process for a
  typical Shack-Hartmann wavefront sensor.
\item A schematic diagram showing the parallelization of the
  wavefront sensor pipeline within the FPGA.  All stages operate
  simultaneously.  Here, NOP means no operation is computed, i.e.\ the
  input data is not valid.
\item A figure showing the performance increase when using an FPGA
  instead of a CPU for wavefront sensing pipeline algorithms, as a
  function of the FPGA clock rate.  The dotted line above 170~MHz
  shows the speeds at which the \fpga algorithm becomes unreliable due
  to overclocking.  Here, the centroid locations of 1024
  Shack-Hartmann sub-apertures have been computed at each \fpga clock
  frequency.
\item A figure showing the computation time for the wavefront
  sensing pipeline in hardware (solid curve) and software (dotted
  curve) as a function of the number of sub-apertures to be evaluated.
\item A figure showing the ratio of CPU to FPGA computation time
  for the wavefront sensing pipeline as a function of the number of
  sub-apertures to be evaluated.  A typical current AO system will
  contain 100 sub-apertures, while future AO systems are designed with
  over 100,000 sub-apertures.  The number of image integrations
  carried out before the CCD readout is simulated is 1 (solid curve),
  2 (dotted curve), 4 (dashed curve) and 8 (dot-dashed curve).
\item A figure showing the ratio of CPU to FPGA computation time
  for the wavefront sensing pipeline as a function of the number of
  pupil phase values per sub-aperture (first number for each bar) and
  the number of simulated CCD pixels (in each dimension) per
  sub-aperture (second number for each bar). 
\end{enumerate}

\end{document}